# New phase space of hardness materials and synergic enhancement of hardness and toughness in superconducting Ti$_2$Co and Ti$_4$Co$_2$X (X = B, C, N, O)

Lifen Shi[1,2,3 =], Keyuan Ma[3,4 =], Jingyu Hou[5 =], Pan Ying[5 =], Ningning Wang[1,2], Xiaojun Xiang[1,6], Pengtao Yang[1,2], Xiaohui Yu[1,2], Huiyang Gou[7], Jianping Sun[1,2], Yoshiya Uwatoko[8], Fabian O. von Rohr[4,9*], Xiang-Feng Zhou[5*], Bosen Wang[1,2*] and Jinguang Cheng[1,2*]

[1]*Beijing National Laboratory for Condensed Matter Physics and Institute of Physics, Chinese Academy of Sciences, Beijing 100190, China*

[2]*School of Physical Sciences, University of Chinese Academy of Sciences, Beijing 100190, China*

[3]*Max-Planck-Institute for Chemical Physics of Solids, 01187 Dresden, Germany*

[4]*Department of Chemistry, University of Zurich, CH-8057 Zurich, Switzerland*

[5]*Center for High Pressure Science, State Key Laboratory of Metastable Materials Science and Technology, School of Science, Yanshan University, Qinhuangdao 066004, China*

[6]*Institute of Atomic and Molecular Physics, Sichuan University, Chengdu 61006S, China*

[7]*Center for High Pressure Science and Technology Advanced Research Xibeiwang East Rd. Haidian, Beijing 100094, China*

[8]*Institute for Solid State Physics, University of Tokyo, Kashiwanoha 5-1-5, Kashiwa, Chiba 277-8581, Japan*

[9]*Department of Quantum Matter Physics, University of Geneva, CH-1211 Geneva, Switzerland*

= These authors contributed equally to this work.

*Corresponding authors: bswang@iphy.ac.cn (BSW); Fabian.VonRohr@unige.ch (FBV); xfzhou@ysu.edu.cn (ZXF); jgcheng@iphy.ac.cn (JGC)

## Abstract

Compared to traditional superhard materials with high electron density and strong covalent bonds, alloy materials mainly composed of metallic bonding structures typically have great toughness and lower hardness. Breaking through the limits of alloy materials is a preface and long-term topic, which is of great significance and value for improving the comprehensive mechanical properties of alloy materials. Here, we report on the discovery of a cubic alloy semiconducting material $Ti_2Co$ with large Vickers of hardness $H_v^{exp}$ ~ 6.7 GPa and low fracture toughness of $K_{IC}^{exp}$ ~ 1.51 MPa·m$^{1/2}$. Unexpectedly, the former value is nearly triple of the $H_v^{cal}$ ~ 2.66 GPa predicted by density functional theory (DFT) calculations and the latter value is about one or two orders of magnitude smaller than that of ordinary titanium alloy materials ($K_{IC}^{exp}$ ~ 30-120 MPa·m$^{1/2}$). These specifications place $Ti_2Co$ far from the phase space of the known alloy materials, but close to medium hardness materials such as MgO or $TiO_2$. Upon incorporation of oxygen into structural void positions, both values were simultaneously improved for $Ti_4Co_2O$ to ~ 9.7 GPa and ~ 2.19 MPa·m$^{1/2}$, respectively. Further DFT calculations on the electron localization function of $Ti_4Co_2X$ ($X$ = B, C, N, O) vs. the interstitial elements indicate that these simultaneous improvements originate from the coexistence of Ti-Co metallic bonds, the emergence of newly oriented Ti-$X$ covalent bonds, and the increase of electron concentration. Moreover, the large difference between $H_v^{exp}$ and $H_v^{cal}$ of $Ti_2Co$ suggests underlying mechanism concerning the absence of the O(16$d$) or Ti2-O bonds in the O-(Ti2)$_6$ octahedron. Our discovery expands the phase space of alloy materials and illuminates the path of exploring superconducting materials with excellent mechanical performances.

# Introduction

Titanium and its alloy materials are extensively applied in aerospace chemical, petrochemical and marine industries as structure materials due to their low density, high specific strength and excellent corrosion resistance[1-4]. They are also one of the most attractive metallic biomedical materials because of their outstanding biocompatibility, bio-corrosion resistance, absence of tissue toxicity and allergic reactions, and high strength and low elastic modulus compatible to natural human bones[3]. However, high price, poor toughness, poor wear resistance, and poor high-temperature oxidation resistance are currently the main obstacles that restrict titanium alloys from cutting-edge industrial field applications, especially in wear-resistant materials[5]. Therefore, it is urgent and valuable to explore and optimize new titanium alloy materials with excellent functionalities and propose potential methods of performance optimization.

Recently, the cubic $\eta$-carbide compounds $Ti_4M_2O$ ($M$ = Co, Rh and Ir) have been identified as novel superconductors with large upper critical field $B_{c2}(0)$. Particularly noteworthy in this regard is the $B_{c2}(0)$ of $Ti_4Co_2O$ and $Ti_4Ir_2O$, which both exceed the weak-coupling Bardeen-Cooper-Schrieffer (BCS) Pauli limit $B_p^{BCS}$ ($\approx 1.84[T/K]T_c$)[6, 7]. Some members of cubic $\eta$-carbide compounds have been discovered having larger bulk modulus of $\sim$ 252 GPa ($Ti_4Ir_2O$)[8] and $\sim$ 278.6 GPa ($Nb_4Rh_2C_{1-\delta}$). This indicates that they may have higher hardness or excellent mechanical properties, but there have been no reports so far, mainly because of the difficulty in growing high-quality single crystals. At the same time, the intermetallic compound $Ti_2Co$, isostructural to the cubic $\eta$-carbide compounds $Ti_4M_2O$, is widely used in the preparation of surface coatings on titanium alloys due to its simple structure, high wear resistance, and good compatibility with titanium metals[9-11]. For example, toughness and wear resistance can been significantly improved by preparing $Ti_2Co$/TiCo dual phase alloy coatings with TiCo as the substrate[9] or TiC/Ti-$Ti_2Co$ coatings with Ti-$Ti_2Co$ dual phase structure as the substrate on the surface of titanium alloys[11]. However, due to the small Vickers hardness and low melting point, $Ti_2Co$ alloy materials are still difficult to meet the necessary demands under harsh conditions. Therefore, improving the comprehensive properties of titanium alloys such as hardness, toughness, and damage tolerance has always been a key scientific issue that needs to be addressed.

Previous density functional theory (DFT) calculations of $Ti_4M_2O$ have revealed that the 3$d$-bands of Ti dominate the density of state at the Fermi level, and the light element O-2$p$ bands are far below the Fermi level and just play the role in stabilizing crystal structure and hole doping[6, 7]. On one hand, the relatively narrow $d$ bands with sufficient electronic correlation make it much different from ordinary alloy materials as well as superhard materials with high electron density. On the other hand, in traditional superhard materials, the short and strong covalent bonds of light elements are critical ingredients apart from alloy materials dominated by metallic bonding structures. In the present Ti$_4M_2$O, how the newly formed Ti-O bonds affect the electrical properties including superconducting transition, and mechanical properties of systems is an important issue worth studying. It may provide a bridge between them to fully understand their underlying physical mechanisms. For example, the coexisting metallic and covalent bonds in a single-phase material may provide new perspectives and physical mechanisms for comprehensive understanding on the mechanical properties of alloy materials.

Here we have grown high-quality single crystals of $\eta$-carbide compounds Ti$_2$Co and Ti$_4$Co$_2$O and discovered the simultaneous improvement of Vickers hardness (from ~ 6.7 GPa in Ti$_2$Co to ~ 9.7 GPa in Ti$_4$Co$_2$O), and fracture toughness (~ 1.51 MPa·m$^{1/2}$ in Ti$_2$Co increased by nearly 1.5 times to ~ 2.19 MPa·m$^{1/2}$). For Ti$_2$Co, the $H_v^{exp}$ ~ 6.7 GPa is nearly triply beyond the predicted value of $H_v^{cal}$ ~ 2.66 GPa by our initial DFT calculations and the $K_{IC}^{exp}$ ~ 1.51 MPa·m$^{1/2}$ is about one or two orders smaller than that of ordinary titanium alloy materials. The subsequent DFT calculations on the electron localization function of Ti$_4$Co$_2X$ ($X$ = B, C, N, O) vs. the interstitial light elements strongly indicate that the coexistence of Ti-Co metallic bonds and the newly emerged Ti-$X$ ($X$ = B, C, N, O) covalent bonds are responsible for the enhancement of mechanical properties. However, the failure of hardness prediction by DFT for Ti$_2$Co may indicate a new mechanism concerning the absence of the O(16$d$) or Ti2-O bonds in the O-(Ti2)$_6$ octahedron. This discovery not only broadens the phase space of alloy materials but shows great potential for Ti$_4$Co$_2X$ to explore practical superconducting materials with excellent mechanical properties.

## II. Experimental methods

Single crystals of Ti$_2$Co and Ti$_4$Co$_2$O were prepared from titanium powder (99.9%, Alfa Aesar), cobalt powder (99.99%, Strem Chemicals), and titanium dioxide powder (purity: 99.9%, Sigma-Aldrich) by arc melting method. The reactants were weighed and mixed in stoichiometric ratios and pressed into a dense rod (8 mm diameter) with pressure. Then the rods were melted in an arc furnace in a purified argon atmosphere on a water-cooled copper plate. The samples were flipped over and molten 10 times to ensure optimal homogeneity. After arc melting, the samples were sealed in quartz tubes under 1/3 atm argon and annealed in a furnace for 30 days at 1000 °C. Finally, the quartz tubes were cooled down to room temperature by quenching in water. Large crystals of Ti$_2$Co and Ti$_4$Co$_2$O were obtained by mechanical breaking of the annealed samples. Room-temperature XRD was performed by X-ray diffractometer with Cu $K_\alpha$ radiation ($\lambda$ = 0.15406 nm). Electrical resistivity $\rho(T)$ was measured by standard four-probe method on the commercial Physical Property Measurement System (PPMS-9 T, Quantum Design).

The Vickers hardness ($H_v$) was measured with a standard square-pyramidal diamond indenter on the crystallographic plane of single-crystal Ti$_2$Co and Ti$_4$Co$_2$O by using a Micro-Hardness Tester (KB30S). The applied forces change from 0.098 to 9.8 N; both the loading and dwell times were 15 s. The estimated $H_v$ values were determined via the equation $H_v = 1854.4F/d^2$, where the $F$ value is the applied load force in Newtons and the $d$ value is the mean of the two diagonals of the Vickers indention in micrometers. At least six indentations obtained at each load were used to determine the asymptotic hardness value. The radial cracks formed on the crystal plane at the load of 9.8 N were used to determine the Fracture toughness ($K_{IC}$) using the equation $K_{IC} = 0.016(E/H_v)^{0.5} \cdot (F/c^{1.5})$, where the $C$ (in micrometers) is the average length of radial cracks and the $E$ value is the Young's modulus (in gigapascals) of bulk Ti$_2$Co and Ti$_4$Co$_2$O.

The DFT calculations on structural optimization, charge density and elastic constants of Ti$_4$Co$_2$X (X = B, C, N, O) were performed using Vienna Ab initio Simulation Package code[12]. Generalized gradient approximation[13] with Perdew-Burke-Ernzerhof[14] functional was adopted to depict electron exchange-correlation energy. The projector augmented-wave[15] pseudopotentials with 3s$^2$3p$^6$3d$^2$4s$^2$ (Ti), 3s$^2$3p$^6$3d$^7$4s$^2$ (Co), 2s$^2$2p$^1$ (B), 2s$^2$2p$^2$ (C), 2s$^2$2p$^3$ (N) and 2s$^2$2p$^4$ (O) treated as valence electrons are used to describe the electron-ion interaction. A cutoff energy of 550 eV and fine k-point grid of 2π × 0.03 Å$^{-1}$ were chosen to ensure energy and force convergences better than 10$^{-8}$ eV and 10$^{-3}$ eV/Å. The elastic constants are calculated via the stress-strain approach[16].

## III. Results and discussion

Figs. 1(a)-(e) compare the crystal structure of single-crystal Ti$_2$Co and Ti$_4$Co$_2$O. Both crystallize into a cubic structure with space group *Fd*-3*m* (No.227), i.e., the Ti1(16*c*), Ti2(48*f*) and Co(32*e*) atoms form a Ti$_2$Ni-type framework with the interstitial O atoms at the 16*d* site. Accordingly, the individual corner-sharing Ti1 and Co tetrahedra form a "*stella quadrangular*" lattice[17]; the O(16*d*) forms a O-(Ti2)$_6$ octahedron with the six nearest Ti2(48*f*), which has an additional Ti-O bond compared to Ti$_2$Co; and the corner-sharing Ti1-Co tetrahedra and Ti2-O octahedra interweave to form a three-dimensional network structure. Figs. 1(f) and (g) show the room-temperature XRD patterns of $Ti_2Co$ and $Ti_4Co_2O$ after Rietveld refinements using the FULLPROF software. The obtained lattice parameters are $a$ = 11.2781(5) Å for $Ti_2Co$ and $a$ = 11.3025(6) Å for $Ti_4Co_2O$, respectively, which are consistent with the previously reported values[17]. It is note that the volume of $Ti_4Co_2O$ is expanded by only 0.41% compared to that of $Ti_2Co$ after introducing the interstitial O atom. This suggests that the newly emerged Ti2-O covalent bonds in the O-$(Ti2)_6$ octahedron have experienced more or less volume contraction, offsetting the overall volume expansion of $Ti_4Co_2O$[17]. Here, the acquisition of high-quality single crystals ensures the reliability of the following mechanical performances.

Fig. 2(a) present the Vickers hardness ($H_v$) of single-crystal $Ti_2Co$ and $Ti_4Co_2O$ by using a four-sided pyramidal diamond indenter. The variations of $H_v$ vs the applied loading force were recorded. For both crystals, there is a dramatic decline of hardness values with the increase of applied load force. Clearly, there is an asymptotic hardness value ~ 6.7 GPa when the load force is 9.8 N for $Ti_2Co$. For comparison, the $H_v$ ~ 6.7 GPa is almost the upper limit of the current hardness of all the known alloy materials[18, 19]: it is larger than those of conventional alloys, such as Ni-, Ti-, Zr- alloys and the upper limit of stainless steel, and close to those of moderately strong W alloys and some oxide materials, $TiO_2$, MgO, etc[18, 19]. Such a large $H_v$ demonstrates a unique physical mechanism of hardness in $Ti_2Co$ because it seems difficult to fully explain this strange feature only considering weak metallic bonds. Unexpectedly, after embedding into the light element O in $Ti_2Co$, the hardness of $Ti_4Co_2O$ is significantly increased to ~ 9.7 GPa under the same loading force of 9.8 N. We note that the hardness of $Ti_2Co$ and $Ti_4Co_2O$ are close to reported records of all the intermetallic compounds[18, 19]. Such significant growth of Vickers hardness is unexpected, and its underlying mechanism deserves further investigation as described below.

Moreover, the fracture toughness $K_{IC}$ was also measured for both $Ti_2Co$ and $Ti_4Co_2O$ by indentation method and the morphology of the measured crystals are shown in Figs. 2(c)-(d). Both exhibit cracks at the four corners of the indentation under the same loading force of 9.8 N, but the indentation area of $Ti_4Co_2O$ in Fig. 2(d) is relatively smaller and the crack length is shorter compared to those of $Ti_2Co$ in Fig. 2(c). This clearly indicates that $Ti_4Co_2O$ has better hardness and fracture toughness. In detail, the Fracture toughness ($K_{IC}$) was determined by using the equation $K_{IC} = 0.016(E/H_v)^{0.5} \cdot (F/c^{1.5})$, where the $C$ (in micrometers) and the $E$ represents the average length of radial cracks and the Young's modulus (in gigapascals), respectively. The obtained $K_{IC}$ values are ~ 1.51 MPa·m$^{1/2}$ and ~ 2.19 MPa·m$^{1/2}$ for Ti$_2$Co and Ti$_4$Co$_2$O, respectively. By comparison, the $K_{IC}$ ~ 1.51 MPa·m$^{1/2}$ in Ti$_2$Co and ~ 2.19 MPa·m$^{1/2}$ in Ti$_4$Co$_2$O are about one or two orders of magnitude smaller in toughness than typical hardness alloy materials, but close to those of the spinel materials as well as MgO, TiO$_2$, etc[18, 19]. Such a small value of fracture toughness indicates that they are brittle materials, which is unusual and in marked contrast with the ductility characteristics of conventional alloys and more like glass for Ti$_2$Co. Interestingly, the $K_{IC}$ ~ 2.19 MPa·m$^{1/2}$ in Ti$_4$Co$_2$O is nearly 1.5 times that of ~ 1.51 MPa·m$^{1/2}$ in Ti$_2$Co.

Combining these results, the insertion of O in Ti$_2$Co leads to simultaneous enhancement of both hardness and fracture toughness, which is rarely seen in common materials. For comparison, phase space of hardness (GPa) and Fracture toughness (MPa·m$^{1/2}$) is constructed in Fig. 3 for typical superhard materials[18-27], including the Mg-, Cu-, Ni-, Ti-, Zr-, Al-, W-alloys, Steels, WC, ZrO$_2$, Al$_2$O$_3$, Si$_3$N$_4$, SiC, AlN, B$_4$C, TiO$_2$, MgO, spinel, cBN, and diamond. We find that the phase space of Ti$_2$Co and Ti$_4$Co$_2$O is far from that of conventional alloy materials and closer to medium hardness materials such as spinel, MgO, TiO$_2$[18, 19]. This discovery broadens phase space of hardness alloy materials and proposes new methods for simultaneous enhancement of the hardness and toughness of alloys.

In addition to the improvement of mechanical properties by interstitial oxygen, the electrical transport properties are also significantly modified. As shown in Fig. 2(b), Ti$_2$Co exhibits a semimetallic behavior in the temperature range of 2-300 K and undergoes a superconducting transition with the transition temperature of $T_c^{onset}$ ~ 0.8 K; while Ti$_4$Co$_2$O displays a typical metallic behavior in the same temperature range and a higher $T_c$ ~ 2.8 K, which is consistent with earlier reports[6]. Previous DFT calculations have revealed the density of states at Fermi level $N(E_F) \approx 7.99$ states/eV/2f.u. in Ti$_2$Co and $\approx 10.2$ states/eV/f.u. in Ti$_4$Co$_2$O, which are in good agreement with the observed increase of the Sommerfeld coefficient from ~5.0 mJ mol$^{-1}$ K$^{-2}$ in Ti$_2$Co to 39.98 mJ mol$^{-1}$ K$^{-2}$ in Ti$_4$Co$_2$O [6]. Since the interstitial O atoms not only introduce hole charge carriers but also form strong Ti2-O covalent bonds, these factors together should be responsible for the observed concomitant improvements of Vickers hardness, fracture toughness, and superconductivity

To gain a deeper understanding of these features, the DFT calculations were performed to characterize the electron localization function of Ti$_2$Co and Ti$_4$Co$_2$X ($X$ = B, C, N, O) vs. the interstitial $X$ elements, which is widely used to analyze the bonding behavior of systems. In Figs. 4(a)-(b) and S1, the electrons around the Ti and Co atoms transfer to the lattice interval near the $X$ atom sites, and there is a significant charge convergence along the direction of Ti1-$X$ bonding, indicating a metallic bond of Ti1-Co atoms and a Ti2-$X$ covalent interaction. It is well known that strong ionic bonds and weak metallic bonds lack directionality, while strong covalent bonds have well-defined directionality. In this regard, the Ti2-$X$ covalent interaction in Ti$_4$Co$_2$$X$ ($X$ = B, C, N, O) fits perfectly two conditions for designing superhard materials: the high electron concentration and the oriented covalent bonds[18-27].

The elastic constant parameter can describe the response of the lattice to macroscopic stress and plays an important role in the strength of materials. Therefore, the detailed research on it can further verify and deepen our understanding on these experiments. In this work, the elastic constants of Ti$_2$Co and Ti$_4$Co$_2$$X$ ($X$ = B, C, N, O) were calculated through DFT calculations and listed in Table SI. For the cubic phase, only three independent terms, $C_{11}$, $C_{12}$ and $C_{44}$, are considered, representing deformation due to stress along three different axis directions. Our calculations can verify that Ti$_2$Co and Ti$_4$Co$_2$$X$ have stable mechanical structures according to their corresponding stability criterion[28]: $C_{11} > 0$, $C_{12} > 0$, $C_{11} - C_{12} > 0$, $C_{11} + 2C_{12} > 0$. Based on the estimated elastic constants of Ti$_2$Co and Ti$_4$Co$_2$$X$ ($X$ = B, C, N, O), the empirical expressions by the Voigt-Reuss-Hill theory[29-31] was adopted to calculate series parameters including bulk modulus ($B$), Young's modulus ($E$), shear modulus ($G$), and Poisson's ratio ($\sigma$). As shown in Table SII, it can be seen that the values of the $B$, $E$ and $G$ are enhanced when the light elements B, C, N and O are inserted into Ti$_2$Co. Our results imply that Ti$_4$Co$_2$$X$ ($X$ = B, C, N, O) compounds have a relatively good deformation resistance and stronger ability to resist shear strain as well as good rigidity. Meanwhile, the $B/G$[31], the $\sigma$ and the Cauchy pressure $C'$ values are significantly reduced, indicating a deterioration in elasticity and ductility after the interstitial of $X$ elements. This abnormal observation indicates that there may be other key mechanisms that determine the fracture toughness of Ti$_4$Co$_2$$X$. In addition, compared with Ti$_2$Co, the universal elastic anisotropy $A^U$ value of Ti$_4$Co$_2$$X$ ($X$ = B, C, N, O) is significantly reduced[32], which means that the anisotropy characteristics become weaker and may be beneficial to reducing cracks of hardness material in practical applications. Finally, our results confirm that the introduction of light elements can improve the strength, hardness, and stiffness of the Ti$_4$Co$_2$$X$ in Fig. 3 and Table SII.

According to the evaluation formulas of Vickers hardness $H_v^{cal} = 0.92k^{1.137}G^{0.708}$ [33], and fracture toughness $K_{IC}^{cal} = V_0^{1/6} \cdot G \cdot (B/G)^{1/2}$ [34], the estimated values of $H_v^{cal}$ and $K_{IC}^{cal}$ for Ti$_2$Co and Ti$_4$Co$_2$$X$ ($X$ = B, C, N, O) were listed in Table SII. It can be seen that the $H_v^{cal}$ values are ≈ 2.66 and 9.38 GPa for Ti$_2$Co and Ti$_4$Co$_2$O, respectively.

The experimental $H_v^{exp} \sim 6.7$ GPa for Ti$_2$Co is well beyond the predicted $H_v^{cal} \sim 2.66$ GPa through DFT calculations; such significant difference doesn't appear in isostructural Ti$_4$Co$_2$O, i.e., the $H_v^{exp} \sim 9.7$ GPa is very close to $H_v^{cal} \sim 9.38$ GPa. Such differences in the $H_v^{exp}$ and $H_v^{cal}$ of Ti$_2$Co naturally reminds us of underlying new mechanism of superhard alloys. Moreover, our results indicate that the current DFT theoretical prediction of hardness and toughness except for the Ti$_2$Co, as well as the trend of changes caused by the insertion of O, are reliable. In this case, we roughly estimated the fracture toughness and hardness values of Ti$_4$Co$_2$X ($X$ = B, C, N, O) with the interstitial of $X$ atoms, and then compared them with the values of Ti$_2$Co as shown in Table SII. We can conclude that the insertion of light elements $X$ ($X$ = B, C, N, O) simultaneously enhances the hardness and fracture toughness of Ti$_4$Co$_2$X alloys. For example, for $X$ = C and N, the $H_v^{cal}$ enhances nearly two times and up to 11.83 and 11.88 GPa, respectively; and the $K_{IC}^{cal}$ increases to 4.53 and 4.77 MPa·m$^{1/2}$, respectively. We note that these values are nearly one or two order smaller than that of ordinary titanium alloy materials ($K_{IC}^{exp} \sim 30$-$120$ MPa·m$^{1/2}$). Based on these, $H_v^{cal}$ and $K_{IC}^{cal}$ of Ti$_4$Co$_2$X ($X$ = B, C, N, O) were summarized in Fig. 3. Our findings greatly expand new phase space of hardness alloy materials different from the conventional ones. Our discovery may provide a new approach for improving the hardness and mechanical properties of Ti$_2$Co by selective oxidation on its surface.

Finally, we discuss briefly the findings of the present study. At first, we report a new phase space of Vickers hardness-fracture toughness in a cubic alloy Ti$_2$Co and Ti$_4$Co$_2$X ($X$ = B, C, N, O), which is far away from the conventional phase space but closer to the medium hardness materials such as MgO, TiO$_2$ [18-27]. However, Ti$_2$Co is a special case because it is difficult to explain such a large hardness and small fracture toughness only considering the weak metallic bonds. Considering the importance of the Ti2-O bonds in the O-(Ti2)$_6$ octahedron in determining the hardness, it seems that the absence of the O(16$d$) may be one origin, for example, the inevitable existence of non-intrinsic phenomena such as lattice disorder and atomic interstitial fillings in actual materials ultimately leads to theoretical and experimental disagreements. There may also be certain ignored theoretical factors such as the octahedral distortion or instability caused by the absence of O(16$d$). Thus, this discovery will attract more attention and in-depth research on the anomalous phenomenon. Secondly, the comprehensive mechanical and electrical performances were simultaneously improved with the insert of $X$ atoms in Ti$_4$Co$_2$X ($X$ = B, C, N, O). Our DFT calculations indicate that the coexistence of Ti-Co metallic bonds and the oriented Ti-$X$ covalent bonds can explain the simultaneous enhancement of mechanical properties. However, for some correlated electron system, whether such enhancements also involve other mechanisms pertinent to the moderate electronic correlations is currently a key scientific issue to be addressed in future. Another thought-provoking question is why the Ti2-O bond isn't an ionic bond but a covalent one. These all make $Ti_2Co$ and its derivative compounds different from the ordinary Ti-O-based oxides. More theoretical calculations are required to reveal the underlying physical mechanisms. Last but not least, our discovery broadens the phase space of hardness alloy materials and proposes new methods for simultaneously improving mechanical and electrical performances. Our discovery illuminates the path of exploring practical superconducting materials with excellent mechanical properties.

## IV. Conclusion

In summary, we report the discovery of a cubic alloy semiconducting material $Ti_2Co$ with large Vickers of hardness $H_v^{exp}$ ~ 6.7 GPa and low fracture toughness of $K_{IC}^{exp}$ ~ 1.51 MPa·m$^{1/2}$. There is a significant difference between experimental and theoretical calculations of hardness for $Ti_2Co$, suggesting a new mechanism concerning structural instability, i.e., the absence of the Ti2-O bonds in the O-(Ti2)$_6$ octahedron. Our findings make $Ti_2Co$ and $Ti_4Co_2O$ far away from the phase space but closer to the medium hardness materials such as spinel. Moreover, a simultaneous improvement of Vickers hardness, fracture toughness and superconductivity through the interstitial elements in $Ti_4Co_2X$ ($X$ = B, C, N, O) was found. The subsequent DFT calculations on the electron localization function of $Ti_4Co_2X$ ($X$ = B, C, N, O) vs. the interstitial elements elucidate that the coexistence of Ti-Co metallic bonds and the oriented Ti-$X$ covalent bonds are responsible for the concurrent enhancement of mechanical and electrical properties.

## Acknowledgements

This work is supported by the National Key Research and Development Program of China (Grants No. 2024YFA1408400, No. 2023YFA1406100, No. 2023YFA1607400, No. 2022YFA1403800, No. 2022YFA1403203), the National Natural Science Foundation of China (Grants No. 12474055, No. 12404067, No. 12025408, No. U23A6003), the Strategic Priority Research Program of CAS (XDB33000000); the Chinese Academy of Sciences President's International Fellowship Initiative (Grants No. 2024PG0003), and the Outstanding member of Youth Promotion Association of Chinese Academy of Sciences (Grants No. Y2022004). This work was supported by the CAC station of Synergetic Extreme Condition User Facility (SECUF, https://cstr.cn/31123.02.SECUF).

# Figure Captions:

**FIG. 1.** (a) Crystal structures of single-crystal $Ti_2Co$ and $Ti_4Co_2O$. (b) Concentrically nested tetrahedra with corner-shared connections composed of Ti1 and Co atoms in $Ti_2Co$ and (c) the basic unit of concentrically nested tetrahedron. (d) Corner-shared network of O-$(Ti2)_6$ octahedra and (e) basic unit of O-$(Ti2)_6$ octahedron in $Ti_4Co_2O$. Observed (open circle), calculated (solid line), and difference (bottom line) XRD profiles of (f) $Ti_2Co$ and (g) $Ti_4Co_2O$ powder sample after Rietveld refinement. The Bragg positions are shown by vertical marks.

**FIG. 2.** (a) The Vickers hardness ($H_v$) as a function of applied loading force from 0.098 N to 9.8 N using a four-sided pyramidal diamond indenter. (b) Temperature dependence of the electrical transport of $Ti_2Co$ and $Ti_4Co_2O$, respectively. The inset shows the enlarged low-temperature data and the arrow indicates superconducting transition. The cracking around Vickers indentations of $Ti_2Co$ and $Ti_4Co_2O$ under a load force of 9.8 N: (c) $Ti_2Co$ and (d) $Ti_4Co_2O$, respectively.

**FIG. 3.** Phase diagram of hardness (GPa) and fracture toughness (MPa·m$^{1/2}$) for various super-hard materials, including the Mg-, Cu-, Ni-, Ti-, Zr-, Al-, W-alloys, Steels, WC, $ZrO_2$, $Al_2O_3$, $Si_3N_4$, SiC, AlN, $B_4C$, $TiO_2$, MgO, spinel, cBN, diamond. The red and blue star represent the $Ti_2Co$ and $Ti_4Co_2O$, and other colors rhombus represent $Ti_4Co_2X$ ($X$ = B, C, N)[18, 19, 21-27], respectively.

**FIG. 4.** DFT calculations for electron localization function map of (110) plane of (a) $Ti_2Co$ and (b) $Ti_4Co_2O$, respectively.

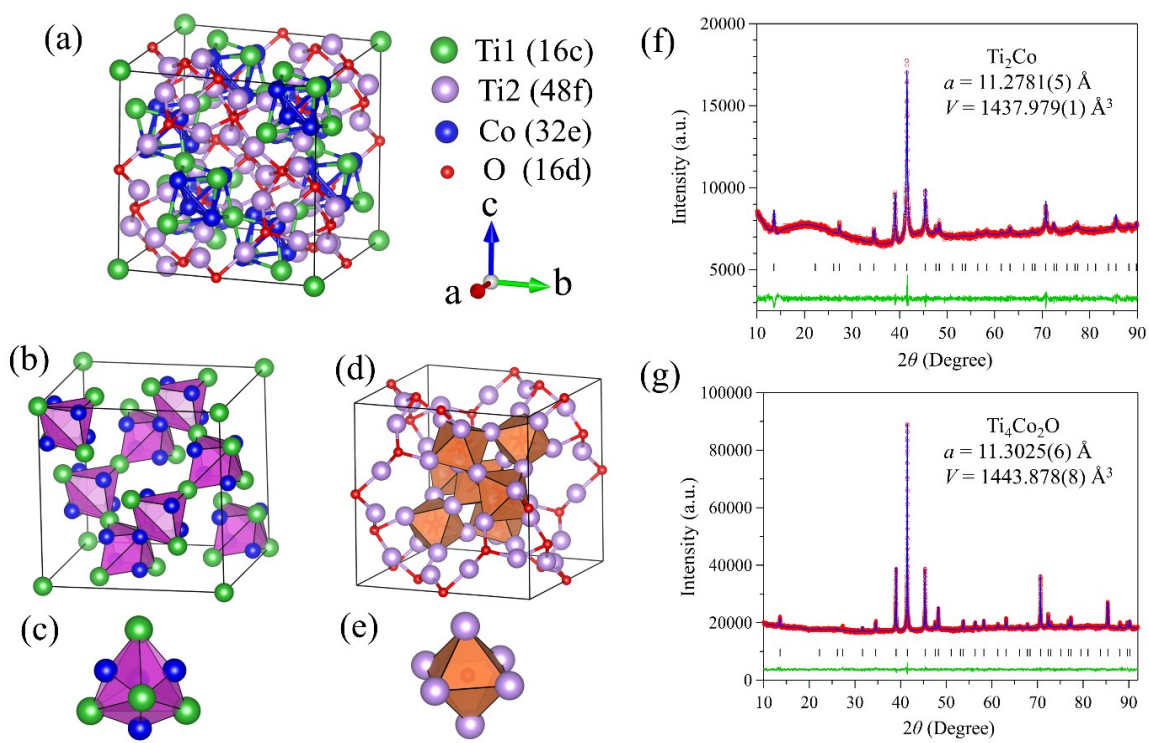

**Figure 2**

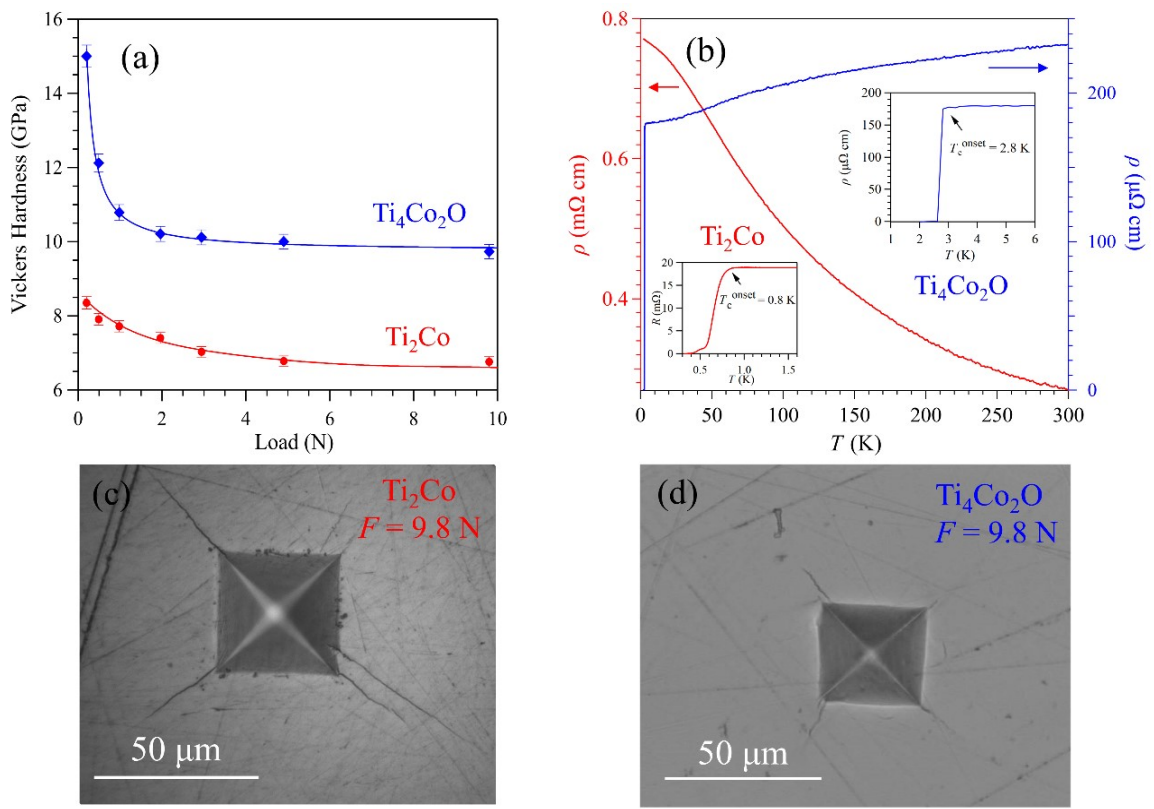

**Figure 3**

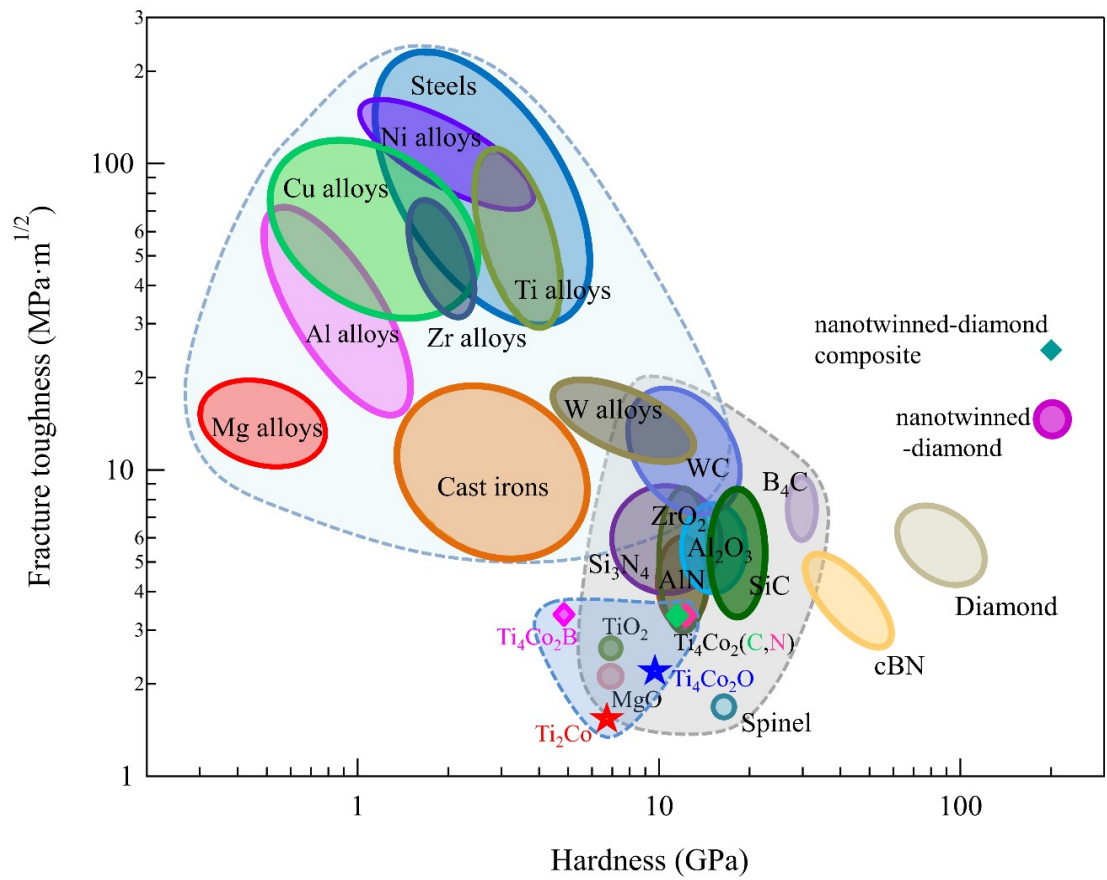

**Figure 4**

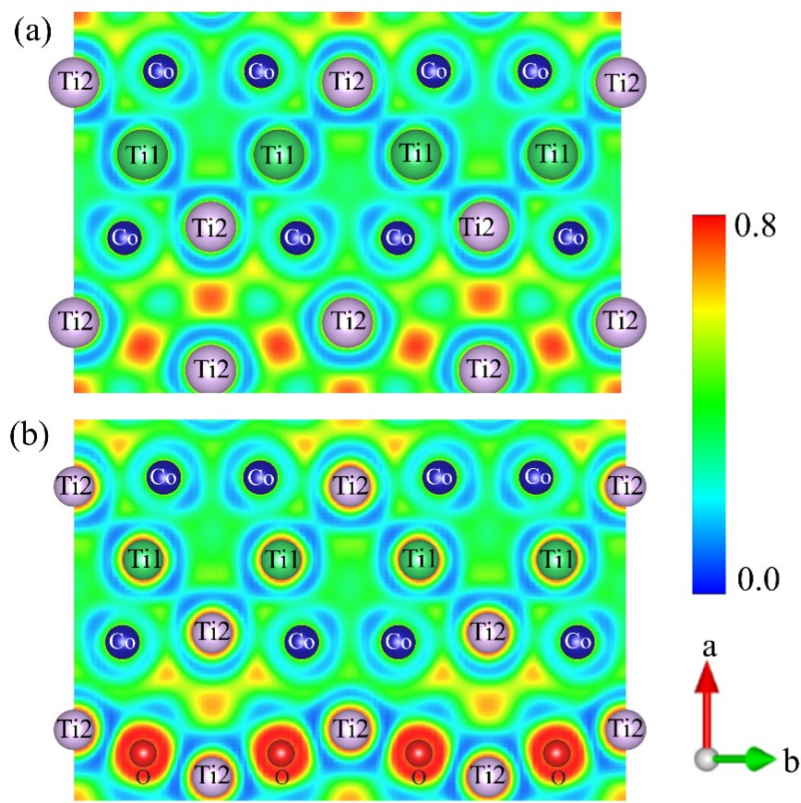